# Evolution of electronic and magnetic properties in Mn- and Co-alloyed ferromagnetic kagome metal $Fe_3Sn_2$

Prajwal M. Laxmeesha[†,1], Rajesh Dutta[†,1], Rajeev Kumar Rai[2], Sharup Sheikh[3], Michael F. DiScala[4], Uditha M. Jayathilake[3], Alexander Velič[1], Tarush Tandon[1], Tessa D. Tucker[1], Christoph Klewe[5], Haile Ambaye[6], Timothy Charlton[6], Tien-Lin Lee[7], Eric A. Stach[2], Kemp W. Plumb[4], Alexander X. Gray[3], Steven J. May[*,1]

[1] *Department of Materials Science and Engineering, Drexel University, Philadelphia, PA, 19104, USA*
[2] *Department of Materials Science and Engineering, University of Pennsylvania, Philadelphia, PA, 19104, USA*
[3] *Department of Physics, Temple University, Philadelphia, PA, 19122, USA*
[4] *Department of Physics, Brown University, Providence, RI, USA 02912, USA*
[5] *Advanced Light Source, Lawrence Berkeley National Laboratory, Berkeley, CA, 94720, USA*
[6] *Neutron Sciences Division, Oak Ridge National Laboratory, Oak Ridge, TN, 37830, USA*
[7] *Diamond Light Source Ltd., Oxfordshire, OX11 0DE, United Kingdom*

† authors contributed equally
* email: smay@drexel.edu

## ABSTRACT

Kagome metals are an intriguing class of quantum materials as the presence of both flat bands and Dirac points provides access to functional properties present in strongly correlated and topological materials. To fully harness these electronic features, the ability to tune the Fermi level relative to the band positions is needed. Here we explore the structural, electronic and magnetic impacts of substitutional alloying within ferromagnetic kagome metal $Fe_3Sn_2$ in thin films grown by molecular beam epitaxy. Transition metals Mn and Co are chosen as substitutes for Fe to reduce or increase the *d*-band electron count, thereby moving the Fermi level accordingly. We find that Co is not incorporated into the $Fe_3Sn_2$ structure but instead results in a two-phase Fe-Co and (Fe,Co)Sn composite. In contrast, $Fe_{3-x}Mn_xSn_2$ films are realized with $x$ up to 1.0, retaining crystalline quality comparable to the parent phase. The incorporation of Mn repositions the flat bands relative to the Fermi level in a manner consistent with hole-doping, as revealed by hard x-ray photoemission and density functional theory. The $Fe_{3-x}Mn_xSn_2$ films retain room temperature ferromagnetism, with x-ray magnetic circular dichroism measurements confirming that the Fe and Mn moments are ferromagnetically aligned. The ability to hole-dope this magnetic kagome metal provides a platform for tuning properties such as anomalous Hall and Nernst responses.

# I. INTRODUCTION

Materials in which transitional metals form a two-dimensional kagome sublattice such as $Fe_3Sn_2$, $TbMn_6Sn_6$ and $KV_3Sb_5$, have received significant interest over the last decade owing to the prominence of flat bands within these materials.[1-7] The strong electron-electron interactions that arise from the flat bands in kagome materials lead to numerous collectively ordered states such as superconductivity, charge density waves, and magnetism.[8-13] In magnetic kagome materials, the position of the Fermi level relative to the flat band can be an important factor in the stability of magnetic order. For instance, CoSn is a non-magnetic kagome metal in which the Fermi level is approximately 0.4 eV above the flat band. Substitution of In for Sn moves the Fermi level 0.1 eV closer to the flat band, resulting in the emergence of antiferromagnetic order.[14] In antiferromagnetic FeSn, moving the Fermi level through alloying on the Fe site with either Co (electron doping) or Mn (hole doping) results in a similar suppression of the Néel temperature at up to 15% substitution concentration,[15, 16] suggesting that simple band filling is not the sole factor dictating magnetic exchange energies. Kagome ferromagnets, such as $Fe_3Sn_2$ and $Co_3Sn_2S_2$, have been shown to host large anomalous Hall and Nernst conductivities,[17-24] however, to date little has been reported on the impact of electron or hole doping on their functional properties or if parent kagome structures are stable when alloying is attempted.[25, 26] Furthermore, as alloying involves higher concentrations of substituted elements compared to conventional semiconductor doping, the substituted element becomes an integral part of the crystal lattice, which may alter electronic band structure beyond a simple movement of the Fermi level.

In this work, we report on the synthesis and magnetic properties of $Fe_3Sn_2$ films that have been alloyed with Mn and Co, motivated by the desire to tune the Fermi level by altering the 3*d*-electron count. $Fe_3Sn_2$ crystallizes in a rhombohedral $R\overline{3}m$ crystal structure, with lattice parameters $a$ = 5.297 Å and $c$ = 19.84 Å.[27] The structure contains a stacking sequence where a single layer of stanene ($Sn_2$) is sandwiched between two layers of $Fe_3Sn$, in which the Fe atoms form a kagome sublattice. The adjacent $Fe_3Sn$ layers are ferromagnetically coupled, leading to a bulk Curie temperature between 640 K and 660 K.[28-30] $Mn_3Sn_2$ crystallizes in the orthorhombic *Pnma* structure and is paramagnetic at room temperature. It orders ferromagnetically below 262 K, while a ferro- to antiferromagnetic transition is observed at 192 K.[31] $Co_3Sn_2$ crystallizes in two distinct structures: a high temperature hexagonal *P6/mmm* structure above ~570 K, and an orthorhombic *Pnma* structure below 570 K.[32] It is ferromagnetic at room temperature.[33] While the transition metal atoms in $Fe_3Sn_2$ arrange themselves in a kagome sublattice, in both $Mn_3Sn_2$ and $Co_3Sn_2$, the kagome atomic arrangement is absent. Both $Mn_3Sn_2$ and $Co_3Sn_2$ have been synthesized and characterized but we are unaware of previous experimental reports on the ternary compounds $Fe_{3-x}Mn_xSn_2$ and $Fe_{3-x}Co_xSn_2$. In a previous computational study of hole and electron doping of $Fe_3Sn_2$, Adams and coauthors used density functional theory (DFT) to show that by substituting Mn for Fe in $Fe_{3-x}Mn_xSn_2$ (hole doping), the Dirac points initially disappear at $x = 0.5$ but reappear and shift closer to the Fermi level at a stoichiometry of $x = 1$.[34] Contrarily, Co substitution in $Fe_{3-x}Co_xSn_2$ (electron doping) led to the disappearance of Weyl points from the band structure at both $x = 0.5$ and $x = 1.0$. While these computational results highlight electronic changes that may be induced with metal alloying, experimental work is clearly needed to understand the phase stability of Mn and Co-alloyed $Fe_3Sn_2$, determine the structure of these materials, and understand how electronic structure and magnetism are impacted by alloying.

Here we report the growth of kagome ferromagnet $Fe_3Sn_2$ films using molecular beam epitaxy (MBE) and the effects of alloying with Mn and Co. We find that Co does not incorporate into a solid solution with $Fe_3Sn_2$ but instead leads to phase separation into a (Fe,Co)Sn kagome phase and a (Fe,Co)-rich phase. In contrast, Mn is incorporated into $Fe_3Sn_2$, while maintaining the $R\bar{3}m$ structure. We have studied the electronic structure via hard x-ray photoemission spectroscopy (HAXPES) measurements[35] and DFT calculations demonstrating that the Fermi level shifts downward, closer to the flat bands with Mn incorporation, confirming that hole-doping is a means for electronic control in this kagome ferromagnet. Concurrently, $Fe_{3-x}Mn_xSn_2$ retains room temperature ferromagnetic order up to $x = 1$, as confirmed using a combination of magnetometry, polarized neutron reflectometry, DFT calculations and x-ray magnetic circular dichroism. Our results indicate that Mn alloying is a viable strategy for tuning the novel functions of the kagome metal $Fe_3Sn_2$ that arise from flat bands and topological features within its band structure such as anomalous or topological Hall effects,[17, 19, 36] Nernst effects,[37] and its use as a ferromagnet in spin-orbit torque structures.[38]

## II. METHODS

### A. Material Synthesis

Epitaxial $Fe_{3-x}Mn_xSn_2$ films and phase-separated (Fe,Co)Sn:(Fe,Co) films were grown using molecular beam epitaxy (Omicron modified LAB-10 system, with base pressure ~$2 \times 10^{-10}$

Torr) on (0001)-oriented $Al_2O_3$ wafers (MTI Corp.). The depositions were performed *via* sublimation of Fe (99.95%, slug, Alfa Aesar), Mn (99.95%, slug, Alfa Aesar) and Co (99.95%, shots, Alfa Aesar) and evaporation of Sn (99.99%, shots, Alfa Aesar) from effusion cells (MBE Komponenten). For growth of the buffer layers, the Co cell was held at ~ 1295°C. For deposition of kagome layers, the Fe and Sn sources were maintained at ~1145°C (±10°C) and ~1020°C, respectively, and the Mn and Co cells were operated at 700°C (±20°C) and 1210°C (±20°C), respectively. All films were grown using co-deposition with continuous fluxes. Shutter times and flux rates were set to control composition and thickness across all films. All films were capped with a thin layer (~4 nm) of insulating $CaF_2$ to mitigate surface oxidation of the kagome films; the $CaF_2$ was evaporated congruently within the MBE from pieces of $CaF_2$ crystals. A quartz crystal microbalance was used to measure the atomic fluxes immediately prior to deposition; the quartz crystal measurements were calibrated using Rutherford backscattering spectrometry (RBS) and energy dispersive x-ray spectroscopy. Prior to deposition, the substrates were sonicated in an acetone bath for 15 minutes, loaded into the chamber and heated to a temperature of ~ 425°C pre-deposition. The Co buffer layer was deposited on $Al_2O_3$ (0001) substrates at ~450°C, then cooled immediately to room temperature. Next, the Fe, Mn/Co and Sn cell temperatures were adjusted to obtain the desired stoichiometry and the $Fe_{3-x}M_xSn_2$ layer deposition was carried out at ~450°C. After kagome deposition, the sample was cooled gradually to room temperature while deposition of $CaF_2$ took place. *In situ* reflection high-energy electron diffraction (RHEED) with an operating voltage of 14.5 kV was used to monitor film quality.

**B. X-ray Spectroscopy, Neutron Reflectometry and Lab-based Measurements**

X-ray diffraction (XRD) and reflectivity (XRR) were measured using a Rigaku SmartLab diffractometer with a Ge (220) double bounced monochromator and Cu $K_{\alpha 1}$ radiation ($\lambda$ = 1.5406 Å). Analysis of x-ray scattering data was carried out using GenX 3.6 software program to quantify lattice parameters, thickness, and interface/surface roughness. RBS was performed at the Laboratory for Surface Modification at Rutgers University and analyzed using the SIMNRA software package. High-angle annular dark-field (HAADF)-scanning transmission electron microscopy (STEM) with EDS was performed using an aberration-corrected JEOL NEOARM operating at 200 kV. The images and spectra were recorded using a 4 cm camera length and 25 mrad convergence angle. Transport measurements and magnetometry were carried out using a Physical Properties Measurement System (Quantum Design) with a vibrating sample magnetometry (VSM) attachment. Polarized neutron reflectometry was performed at MAGREF beamline at the Spallation Neutron Source, Oak Ridge National Laboratory, and analyzed using Refl1D software package. X-ray absorption spectroscopy (XAS) and XMCD were performed at Beamline 4.0.2 at the Advanced Light Source, Lawrence Berkeley National Laboratory. All the XAS-XMCD data were analyzed using sum rules and open-source code xaspy.[39] Bulk-sensitive HAXPES measurements of the valence-band and core-level spectra were carried out with a photon energy of 6.45 keV at Surface and Interface Structural Analysis beamline I09 of the Diamond Light Source.[40] The pass energy of the electron analyzer was set to 200 eV, and total energy resolution of the spectrometer was estimated to be approximately 300 meV. Using a high-resolution Fermi-edge measurement on a standard Au sample, binding energy calibration was performed. To validate the measured intensities, we compared them with theoretical predictions generated using the SESSA simulation software,[41] which accounts for physical parameters like electron inelastic scattering, angular distributions of photoemission (asymmetry parameters), and the overall experimental setup to accurately model the expected core-level signal strengths. All HAXPES measurements were conducted at room temperature under ultra-high vacuum conditions.

### C. DFT calculations and simulation of VB spectra

Spin polarized DFT calculations on bulk $Fe_{3-x}Mn_xSn_2$ were carried out using the Abinit software package[42, 43] to determine the underlying electronic band structure, orbital and momentum resolved partial density of states, and the bulk magnetic moments. The details of the calculation parameters are provided in the supplementary materials (section II). Orbital weighted projected density of states (PDOS) from atomic *s*, *p*, *d*, *f* orbitals of Mn, Fe and Sn were also calculated using DFT. The energy scale of computed PDOS is reversed by replacing negative orbital energy to positive binding energy of HAXPES. Photoelectron cross-sectional weights for different atomic subcells are used here from energy depended tabulated *ab initio* calculations[44, 45] and the process is done using the GALORE software package.[46]

## III. RESULTS

### A. Structural properties

Thin film deposition was carried out using MBE to supply atomic fluxes of Fe, Sn, Co, and Mn to $Al_2O_3$ (0001) substrates. The Fe family of kagome metals is prone to grow as 2D islands when synthesized as thin films on insulating oxide substrates.[47-49] To circumvent this issue, we recently demonstrated that epitaxial FeSn can be grown on Fe and Co buffer layers, which enables smooth and continuous formation of the kagome material.[50] Following a similar strategy, here we utilized a thin Co buffer layer to promote the growth of epitaxial and laterally continuous $Fe_{3-x}M_xSn_2$ ($M$ = Co and Mn) films. The hexagonal Co layer grows with a (001)-orientation, providing a symmetry matched buffer for $c$-axis oriented growth of $Fe_3Sn_2$. However, we find no evidence that the kagome film is strained to the Co, consistent with our previous work on epitaxial FeSn films.[50] Following deposition of the kagome film, the heterostructures were capped with a thin $CaF_2$ layer (~4 nm) to help mitigate oxidation. Reflection high energy electron diffraction (RHEED) was used to monitor the surface crystallinity of each layer of the heterostructure [Fig. S1(a-d)]. The RHEED patterns at the end of deposition of the $Fe_{3-x}M_xSn_2$ layers were streaky and were observed to repeat with every 60° in-plane rotation, indicative of sixfold rotational symmetry, which is to be expected from a $c$-axis oriented hexagonal crystal. The RHEED pattern of the $Fe_{3-x}Mn_xSn_2$ surface showed streaky patterns for $0 \leq x \leq 1$, but turned spotty at higher $x$ values.

We performed specular $2\theta$-$\omega$ x-ray diffraction (XRD) scans to probe the structure and crystallinity of the films. Figure 1(a) shows XRD data from the parent $Fe_3Sn_2$ film (in red), with clear $000l$ Bragg peaks from $Fe_3Sn_2$, Co and the $Al_2O_3$ substrate, proving that the films are crystalline and grow along the $c$-axis. The data from $Fe_{3-x}Mn_xSn_2$ (shown in green with different values of $x$) along with targeted $Fe_{3-x}Co_xSn_2$ depositions (shown in blue with different values of $x$) as also presented in Fig. 1(a). The $Fe_{3-x}Mn_xSn_2$ films exhibit XRD patterns very similar to $Fe_3Sn_2$, indicating that the $Fe_{3-x}Mn_xSn_2$ films are isostructural with the parent $Fe_3Sn_2$ phase. With increasing Mn content, the $000l$ Bragg peaks shift to lower angles indicating an increase in $c$-axis lattice parameters. The x-ray reflectivity (XRR) profiles displayed in Fig. 1(b) for pristine $Fe_3Sn_2$ and $Fe_2MnSn_2$ thin films indicate smooth surface and interface quality. The XRD patterns and XRR profiles were simulated using the GenX package to extract the $c$-axis parameters and thickness of the Co buffer and $Fe_{3-x}Mn_xSn_2$ layers.[51] Thickness of the $Fe_{3-x}Mn_xSn_2$ films range between 20 to 30 nm while the Co buffer layers are 6 to 8 nm. Roughness of the Co and $Fe_{3-x}Mn_xSn_2$ top interfaces range from 1.3 to 1.7 nm and 1.5 to 2.2 nm, respectively. The obtained $c$-axis parameters are plotted in Fig. 1(c), along with lattice parameters obtained from DFT using ground state structure relaxation calculations on bulk materials. There is a broad agreement between the measured and calculated $c$-axis evolution, both of which show a lattice expansion as Mn is substituted. This lattice expansion with Mn alloying was also predicted in a previous DFT study of $Fe_{3-x}Mn_xSn_2$.[34] In-plane azimuth $\phi$-scans along a fixed $\chi$ angle were performed to confirm that the films are epitaxial and are shown in Fig. S1(e,f), where six distinct in-plane peaks corresponding to the $2\bar{1}\bar{1}0$ Bragg peaks are present in the scan. They are spaced 60º apart from each other but are also ~30° apart from the $2\bar{1}\bar{1}0$ Bragg peaks of $Al_2O_3$. This relative rotation between the substrate and the film is attributed to a better crystallographic alignment between the $Al_2O_3$ and kagome layers for epitaxial growth.[50]

We performed cross-sectional scanning transmission electron microscopy (STEM) on $Fe_3Sn_2$ and both Co and Mn-alloyed heterostructures to gain a better understanding of their structures. Figure 1(d,e) shows STEM images from $Fe_3Sn_2$ and $Fe_2MnSn_2$ films, respectively, along with an overlaid cartoon of the $Fe_3Sn_2$ crystal structure. Along a viewing direction of $[10\bar{1}0]$ for $Fe_3Sn_2$, the image shows a clear stacking sequence of $Fe_3Sn$(A)/$Fe_3Sn$(A)/$Sn_2$(B)… $Fe_3Sn$(A)/$Fe_3Sn$(A)/$Sn_2$(B). For the $Fe_2MnSn_2$ sample, the viewing direction is $[11\bar{2}0]$ but the stacking sequence is the same as the parent $Fe_3Sn_2$. The combination of STEM images and the XRD confirms the epitaxial nature and the crystal structure of the Mn-alloyed films to be the same as the parent $Fe_3Sn_2$.

In the case of the targeted $Fe_{3-x}Co_xSn_2$, there is a notable difference between the XRD data from the $Fe_3Sn_2$ and the Co-alloyed films. Most notably the 0003 and 0006 Bragg peaks at ~13° and ~27° disappear and a single peak at ~20° appears. However, the peak at ~41°, which would correspond to the 0009 Bragg peak in a $R\bar{3}m$ structure, remains intact. These changes in the XRD pattern point to the Co-alloyed films adopting a different structure from the parent $Fe_3Sn_2$. The peaks at ~20° and ~41°, interestingly, are characteristic structural peaks for FeSn and CoSn kagome structure as the 0001 and 0002 Bragg reflections in $P6/mmm$ space group.[50, 52] The FeSn kagome structure is closely related to the $Fe_3Sn_2$ structure, differing in that FeSn consists of a $Fe_3Sn$/$Sn_2$/$Fe_3Sn$/$Sn_2$ stacking sequence instead of the kagome bilayer $Fe_3Sn$/$Fe_3Sn$/$Sn_2$ sequence found in $Fe_3Sn_2$.

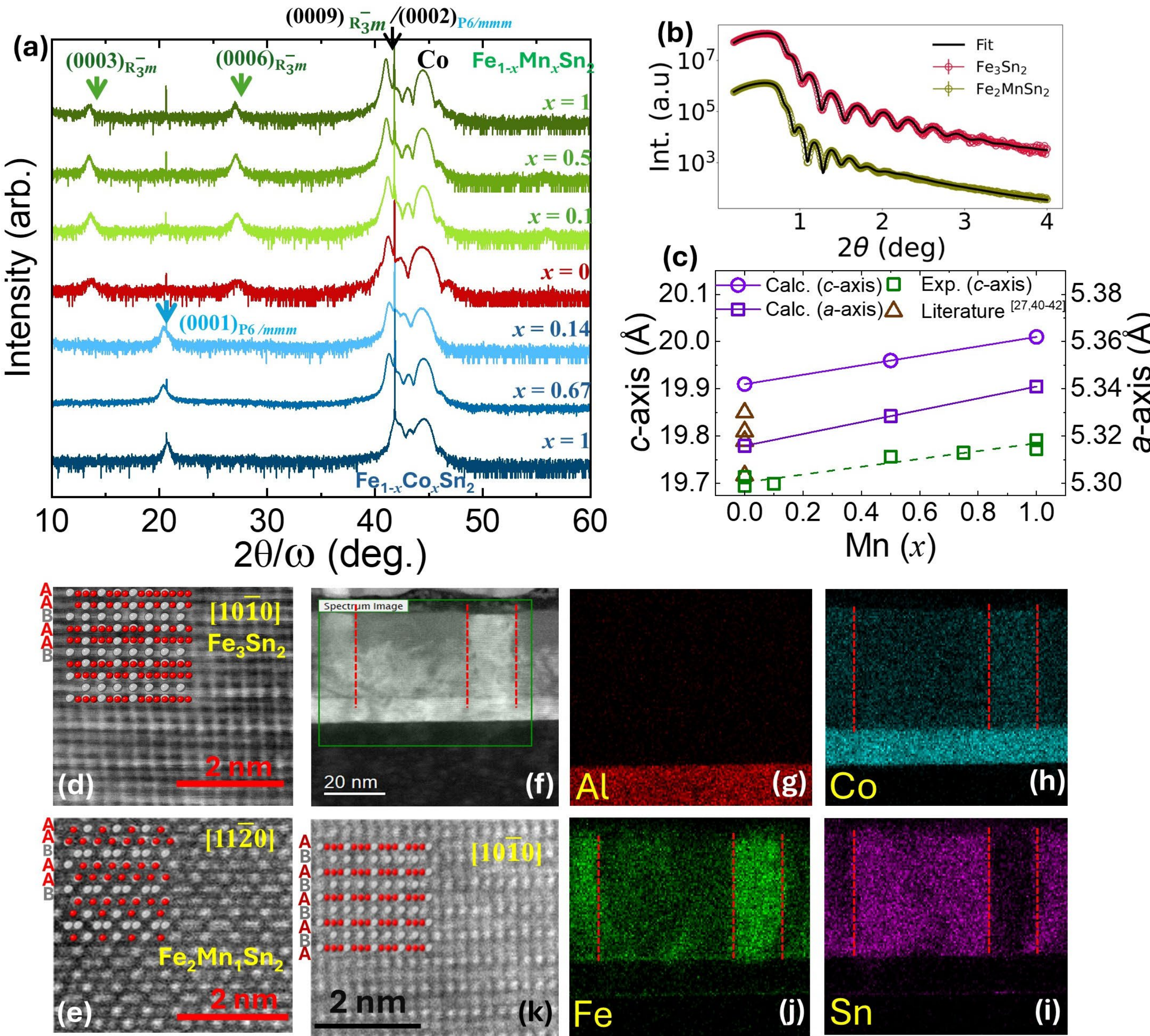

**FIG. 1.** Structural characterization of $Fe_{3-x}M_xSn_2$ ($M$ = Co, Mn) films. (a) XRD $2\theta$-$\omega$ scans from $Fe_{3-x}M_xSn_2$ films with different $x$ concentrations, where data in green (blue) was obtained from Mn-alloyed films (Co-alloyed films). Data from parent $Fe_3Sn_2$ phase is shown in red. (b) XRR data and fits from $Fe_3Sn_2$ and $Fe_2MnSn_2$ thin films. (c) $c$-axis lengths determined from XRD and DFT calculations, as well as those previously reported in the literature.[27, 37, 53, 54] Compositions with multiple points ($x$ = 0.0 and 1.0) are data obtained from separate film depositions demonstrating reproducibility. Cross-sectional STEM image of a (d) $Fe_3Sn_2$ film viewed from the $[10\bar{1}0]$ direction and a (e) $Fe_2Mn_1Sn_2$ film viewed from the $[11\bar{2}0]$ direction. A cartoon inset of the $Fe_3Sn_2$ unit cell is included to visualize the layer stacking. (f) Low-magnification image of a Fe-Co-Sn film showing phase segregated columnar regions, marked by red dotted lines. (g-j) EDS maps corresponding to Al, Co, Sn and Fe respectively. (k) High-magnification image from the kagome region of the Fe-Co-Sn film with an inset of the FeSn crystal structure viewed from the $[10\bar{1}0]$ direction.

In contrast to those of $Fe_{3-x}Mn_xSn_2$, the STEM images of the targeted $Fe_2CoSn_2$ sample reveal lateral non-uniformity and phase separation. The low-magnification image in Figure 1f shows columnar formations within the Fe-Co-Sn film growing along the *c*-axis direction. The boundaries of these columns are marked by red dotted lines. Figure 1(g-j) presents elemental energy-dispersive x-ray spectroscopy (EDS) maps from scans performed over the area marked by the green box in Fig. 1(f). The Fe, Sn and Co maps shows a clear difference in elemental composition between the columnar region and the rest of the film. The Fe and Co maps show higher intensity in the columnar regions than the rest of the films, and contrastingly, the Sn map shows lower intensity in the columnar region than the rest of the film. We attribute this to the formation of two different phases within the sample: a phase that is predominantly Fe-Co within the columnar region and a Fe-Co-Sn alloy in the rest of the film. Figure 1(k) shows a high-magnification image of the Fe-Co-Sn alloy region from a viewing direction of $[10\bar{1}0]$. The stacking sequence can clearly be seen as $Fe_3Sn$(A)/$Sn_2$(B)….$Fe_3Sn$(A)/$Sn_2$(B), which is expected in the FeSn and CoSn-type kagome metals. This structure type is also consistent with the XRD pattern, which suggested the presence of the *P6/mmm* structure (kagome monolayer) in targeted $Fe_{3-x}Co_xSn_2$ films even at $x = 0.1$. We hypothesize that the addition of Co during growth is primarily promoting the formation of ferromagnetic Fe-Co alloys regardless of the concentration, and the remaining Fe and Co atoms bond with Sn to make up (Fe,Co)Sn, in the general reaction: $2Fe + 1Co + 2Sn \rightarrow Fe_yCo_z + 2(Fe_{1-y}Co_{1-z}Sn)$. The Fe-Sn phase diagram supports this scenario, as the $Fe_3Sn_2$ structure is an equilibrium phase only between the narrow temperature range of 610°C to 800°C.[55] Below 610°C, the equilibrium phase behavior at this composition is the formation of a two-phase Fe and FeSn system. While $Fe_3Sn_2$ can be stabilized as an epitaxial film at temperatures below 500°C, as we and others have demonstrated,[37, 54] the addition of Co must further destabilize the $Fe_3Sn_2$ structure, leading to the equilibrium (Fe,Co) and (Fe,Co)Sn mixed phase behavior. In contrast, the enthalpy cost of adding Mn to $Fe_3Sn_2$ appears to not be large enough to destabilize the kagome phase, although future thermodynamic calculations would be useful to quantify the enthalpic differences between Co and Mn substitution in $Fe_3Sn_2$. In our films, the isolated (Fe,Co)Sn regions appear to be epitaxial, as the peaks in the $\phi$-scan (Fig. S1) from the targeted $Fe_2CoSn_2$ film correspond to the FeSn $2\bar{1}\bar{1}0$ Bragg peaks. Given that these films are not a $(Fe,Co)_3Sn_2$ phase, we focus our electronic and magnetic characterization solely on the $Fe_{3-x}Mn_xSn_2$ films.

**B. Electronic structure**

We present the calculated orbital-resolved band structures of ferromagnetic (FM) $Fe_3Sn_2$, $Fe_{2.5}Mn_{0.5}Sn_2$, and $Fe_2MnSn_2$ in Fig. 2(a-c). FM $Fe_3Sn_2$ displays partially flat dispersions of some bands across the large *k*-path [Fig. S2(e)] just below the Fermi level ($E_F$) at ~ -0.13 eV similar to the flat band (FB) found experimentally in $Fe_3Sn_2$ single crystals near -0.2 eV via angle resolved photoemission spectroscopy (ARPES) measurements.[5] The FB is mainly composed of *xy* and *xz* orbital contributions (see Fig. S3 for orbital-resolved DOS in FM and nonmagnetic phase). However, there is another FB composed of mainly *yz* and a small amount of $z^2$ orbital contributions around -0.35 eV, spanning a short region in *k*-space between *Z* and *Γ* that was not observed experimentally. A Dirac cone-like dispersion is present, marked with circle in Fig. 2(a-c), at 0.07 eV below the $E_F$. Additionally, an electron-like pocket can also be observed in the vicinity of the *Γ* point near $E_F$, similar to previous theoretical studies.[34, 56] The Dirac cone-like dispersion and the electron pockets are visible only in the spin majority channel whereas the FBs are from spin minority channel (not shown). Upon Mn-substitution, the FBs and electron pocket move closer to the $E_F$ and the Dirac point moves above $E_F$. At $x = 0.5$, the *xy*/*xz*-derived FB is very close to the $E_F$, making this a particularly promising composition for hosting correlated phenomena. Further increasing the Mn concentration to $x = 1.0$, the FBs coming from the *yz* and $z^2$ orbitals reside just slightly below $E_F$. The movement of these band features with respect to $E_F$ can be seen from total density of states (DOS), which is presented in Fig. 2(d). The shift of the different FBs closer to $E_F$ could have interesting ramifications for the transport and magnetic properties of Mn-alloyed $Fe_3Sn_2$.

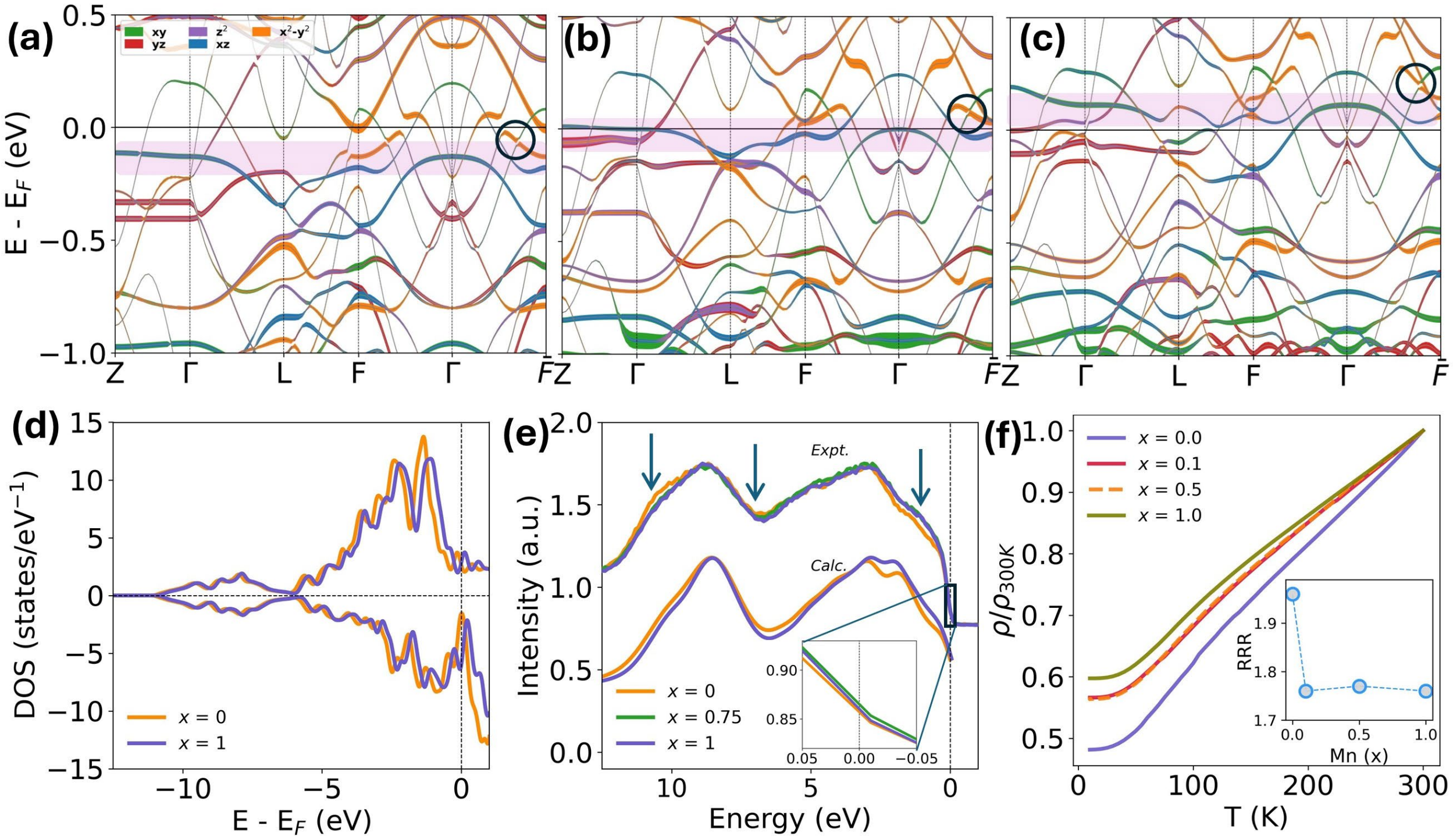

**FIG. 2.** Electronic properties of $Fe_{3-x}Mn_xSn_2$. DFT calculated band structures of (a) $Fe_3Sn_2$, (b) $Fe_{2.5}Mn_{0.5}Sn_2$, and (c) $Fe_2MnSn_2$. The pink rectangular boxes highlight the FB near $E_F$ and the gradual movement of these features closer to the $E_F$ upon Mn alloying. The black circles enclose the Dirac point. (d) The calculated total DOS in the minority and majority spin channel for $x = 0$ and 1. (e) HAXPES measured VB spectra from $Fe_{3-x}Mn_xSn_2$ (for $x = 0$, 0.75, 1) along with the DFT calculated spectra ($x = 0$ and 1.0); experimental data was collected at 300 K. Inset shows a small but clear change in the spectral intensity and spectral shift between $Fe_3Sn_2$ and $Fe_{3-x}Mn_xSn_2$ near $E_F$. (f) Variable temperature resistivity measurement of the heterostructures showing metallic behavior; note that these normalized resistivity data include contributions from both the kagome and underlying Co buffer layer. Inset shows RRR as a function of Mn concentration. The values at $x = 0$ and 1 are taken by averaging the RRR values of several samples of each composition.

Aiming to further probe the band structure near the Fermi level, HAXPES valence band (VB) spectra measurements were performed on $Fe_{3-x}Mn_xSn_2$ ($x = 0$, 0.75, 1) samples. Figure 2(e) shows the experimentally measured and theoretically obtained VB spectra. Near $E_F$, the measured VB spectra show higher intensity for Mn-alloyed films than the parent, as illustrated in the inset. The spectral shifts observed experimentally ($x = 0.75$ and 1.0) is consistent with computationally obtained relative downward movement of $E_F$ of approximately 0.11 and 0.22 eV upon Mn alloying $x = 0.5$ and 1, respectively. Our orbital-weighted DOS from DFT calculations reproduce the features (highlighted with arrows) that are observed in the experimental VB spectra, which points to the accuracy of the DFT calculations. The details of the VB spectra simulation are provided in the supplementary material (section III). In addition, variable temperature resistivity measurements were conducted to confirm the metallic behavior of the heterostructures and normalized resistivity ($\rho/\rho_{300\ K}$) as function of temperature is shown in Fig. 2(f). The figure inset presents the residual resistivity ratio (*RRR*), defined as the ratios of resistivity at 300 K and 10 K as a function of Mn substitution. We attribute the decreased RRR upon Mn substitution to the increased disorder-induced scattering from the random arrangement of Fe/Mn atoms on the kagome sites. Note that these transport results include contributions from both the Co and $Fe_{3-x}Mn_xSn_2$ layers within each heterostructure.

**C. Magnetism**

The $Fe_{3-x}Mn_xSn_2$ films were characterized using dc magnetometry, polarized neutron reflectometry (PNR), and x-ray circular magnetic dichroism (XMCD) to establish how hole-doping alters their magnetic properties. Figure 3(a) shows in-plane magnetic hysteresis loops, normalized by the saturation magnetization, obtained at room temperature from $Fe_3Sn_2$ and Mn-alloyed films of different stoichiometries. All films exhibit ferromagnetism, albeit with a non-monotonic trend in their coercive field ($H_C$). It should be noted that since the Co buffer layer is also ferromagnetic, a signification portion of the magnetic signal is expected to arise from it. To probe and isolate the magnetization contributions from each ferromagnetic layer, we performed room-temperature polarized neutron reflectometry (PNR) on a Co (6 nm) /$Fe_3Sn_2$ (21 nm)/$CaF_2$ (4 nm) film under 0.1 T of external in-plane magnetic field. Figure 3b presents neutron reflectivity profiles up to 0.14 $Å^{-1}$ of momentum transfer ($q$) in the spin up ($R^+$) and spin down ($R^-$) channel. By fitting the PNR data using Refl1D software, we determined the net magnetization in the $Fe_3Sn_2$ to be 1.92 $\mu_B$/Fe, which is in good agreement with the value obtained in $Fe_3Sn_2$ single crystals (1.94 $\mu_B$ at 2 K).[5] In the Co layer, the net magnetization was found to be 1.66 $\mu_B$/Co, which is also close to the bulk Co moment. The inset of Fig. 3(b) shows spin asymmetry, defined as $(R^+-R^-)/(R^++R^-)$, indicating good quality data and model fit. The obtained nuclear and magnetic scattering length density (SLD) profiles are shown in Fig. 3(c). For the best agreement with experimental data, we relaxed the magnetization at the $Fe_3Sn_2/CaF_2$ interface, where the overall magnetization is slightly reduced to ~1.72 $\mu_B$/Fe, which is lower than in the bulk of the film and indicates a possible 2 nm of oxide layer formation. EDS analysis from STEM also confirmed the presence of an oxide layer at the film surface. We also observed a difference between the chemical and magnetic roughness at the Co/$Fe_3Sn_2$ interface to be 2.2 nm and 0.5 nm, respectively. The x-ray SLD profile (not shown) derived from XRR fitting [Fig. 1(b)] performed on the same sample also yielded a similar roughness value of 1.8 nm at the Co/$Fe_3Sn_2$ interface. Given that we find Co to be immiscible in $Fe_3Sn_2$, it is possible that Fe is diffusing into the Co buffer layer forming a magnetic (Co,Fe) phase with a stoichiometry gradient impacting the magnetic profile (magnetic roughness) across the Co/$Fe_3Sn_2$ interface in a manner that doesn't simply scale with the chemical intermixing (chemical roughness).

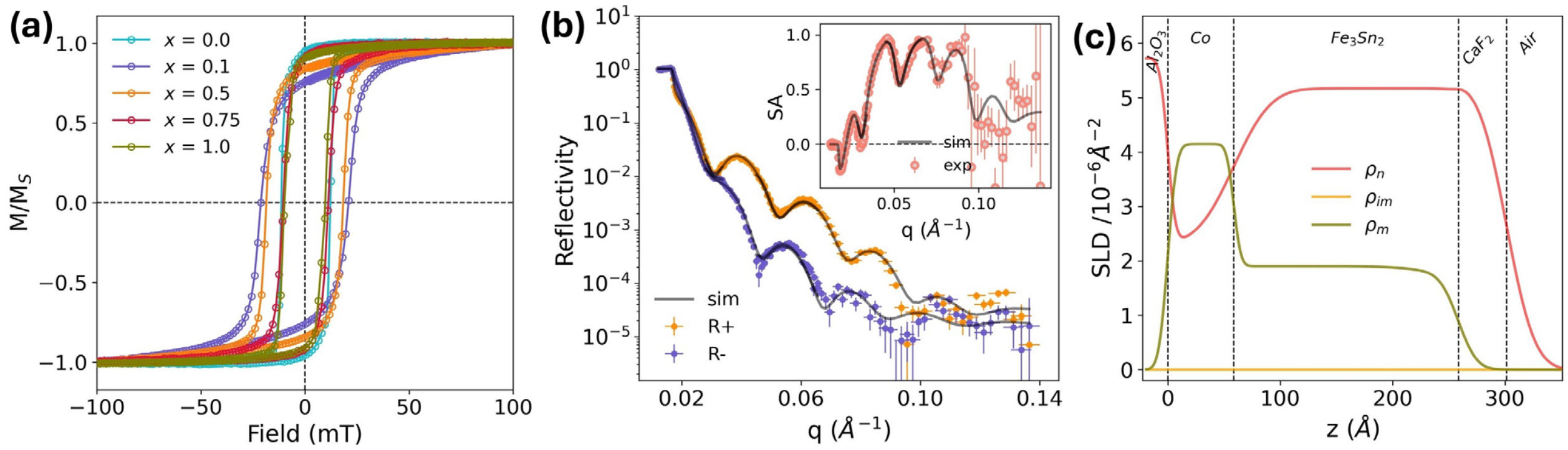

**FIG. 3.** Magnetic properties of $Fe_{3-x}Mn_xSn_2$ films. (a) Normalized *M-H* loops for $x = 0$, 0.1, 0.5, 0.75 and 1, measured at 300 K under in-plane magnetic fields. (b) Measured and simulated PNR data from a Co/$Fe_3Sn_2$ heterostructure at 300 K. Inset shows the spin asymmetry (SA) data along with fit. (c) Magnetic and nuclear scattering length density profiles of the Co/$Fe_3Sn_2$ heterostructure.

Having established the magnitude of net magnetization in the Co buffer layer as well as $Fe_3Sn_2$ layer, we subtracted the Co contribution from the $M_S$ of the whole stack based on the volume of the buffer layer to obtain the magnetic contribution from just the kagome film. Using this method, $M_S$ obtained from magnetometry was calculated to be $1.92 \pm 0.03$ $\mu_B$/Fe atom for the parent $Fe_3Sn_2$ film, which is in excellent agreement with the magnetization obtained from PNR. These values are also consistent with previous reports of $Fe_3Sn_2$.[5, 17, 54] Emboldened by the agreement, we calculated the magnetization in the Mn-alloyed films by similarly subtracting the Co contribution from all $Fe_{3-x}Mn_xSn_2$ films. We find that the addition of Mn leads to a decrease in total moment per formula unit as shown in Fig. 4(c). This figure also presents the net magnetic moment obtained from DFT calculations, which decreases with a much shallower slope compared to the magnetometry data.

In order to extract the elementally resolved spin and orbital moment of Fe and Mn in the alloyed samples, we performed room temperature XMCD measurements under a fixed field of 0.4 T applied along the x-ray incidence angle of 20° with a 90% circularly polarized x-ray beam alternating between left and right circular polarizations. The x-ray absorption spectra were collected in luminescence yield (LY) mode which is a bulk sensitive measurement mode, as the presence of the capping layer and surface oxidation leads to the $L_3$ and $L_2$ edge shoulder peaks, and/or new peaks associated with different nominal Fe and Mn oxidation states. Additionally, a 20° incident angle geometry was utilized because a previous report on $Fe_3Sn_2$ demonstrated that the largest XMCD signal increases with decreasing incident angle relative to the (0001)-surface plane.[57] Figures 4a,b show clear XMCD spectra spanning the Fe $L_{2,3}$-edge and the Mn $L_{2,3}$-edge, respectively, for multiple samples. To quantify the individual contributions of Fe and Mn spin and orbital moment to the overall magnetization of $Fe_{3-x}Mn_xSn_2$, we have carried out a sum rule analysis[58-62], more details of which are given in the Supporting Information along with measured x-ray absorption spectra in Figs. S5 and S6. The net moment per formula unit obtained from XMCD sum rule is provided in Fig. 4(c). To calculate the individual moment (spin and orbital) per transition metal atom, one needs information about the number of holes ($n_H$). Therefore, we have estimated $n_H$ for Fe via matching the known moment of Fe in $Fe_3Sn_2$ from PNR measurement and we obtained $n_H = 3.39$ which is close to the theoretical value.[63] For Mn we have assumed the number of holes to be 4.488.[64] Utilizing these values for $n_H$, sum rule analysis reveals that the Mn alloying in the lattice leads to an initial increase in total moment at $x = 0.1$ and then a decrease with increasing Mn content. The initial increase at $x = 0.1$ comes mainly from an increased Fe moment as can also be seen from the Fe XMCD signal [Fig. 4(a)], which is more intense compared to other samples. The individual spin and orbital moment is displayed in Fig. S8, Supporting Information. The Fe spin moment per hole of parent $Fe_3Sn_2$ (~0.55 $\mu_B$) is consistent with a recent XMCD measurement of bulk $Fe_3Sn_2$ (~0.6 $\mu_B$).[57] We obtain ratios of the orbital to spin moment of 0.08 in the parent $Fe_3Sn_2$ compound, which is significantly less than the value of 0.22 reported in Ref. [57], but is closer to the value (~0.043) for pure iron.[58] In the $Fe_{3-x}Mn_xSn_2$ alloys, we find orbital to spin moment ratios of 0.06 to 0.09 for Fe without any clear dependence on the Mn concentration (Fig. S8).

Accounting for the spin- and orbital-moment contributions from both Fe and Mn, the XMCD results further validate a decrease in net magnetization with increasing Mn concentration.

The decrease in magnetization is less than that measured from magnetometry but more than that obtained from either our DFT calculations or those by Adams et al.[34] That magnetometry results in the lowest magnetization values in Mn alloyed samples could potentially be from non-idealities arising at interfaces where chemical intermixing or a difference in surface oxidation might lead to lower global magnetization. Additional PNR on multiple Mn-alloyed $Fe_3Sn_2$ films, beyond the scope of this work, would provide more insight into the degree of interfacial intermixing and its effect on magnetism, as well as providing information on the possible existence of magnetically suppressed layers near the surface of the $Fe_{3-x}Mn_xSn_2$ films.

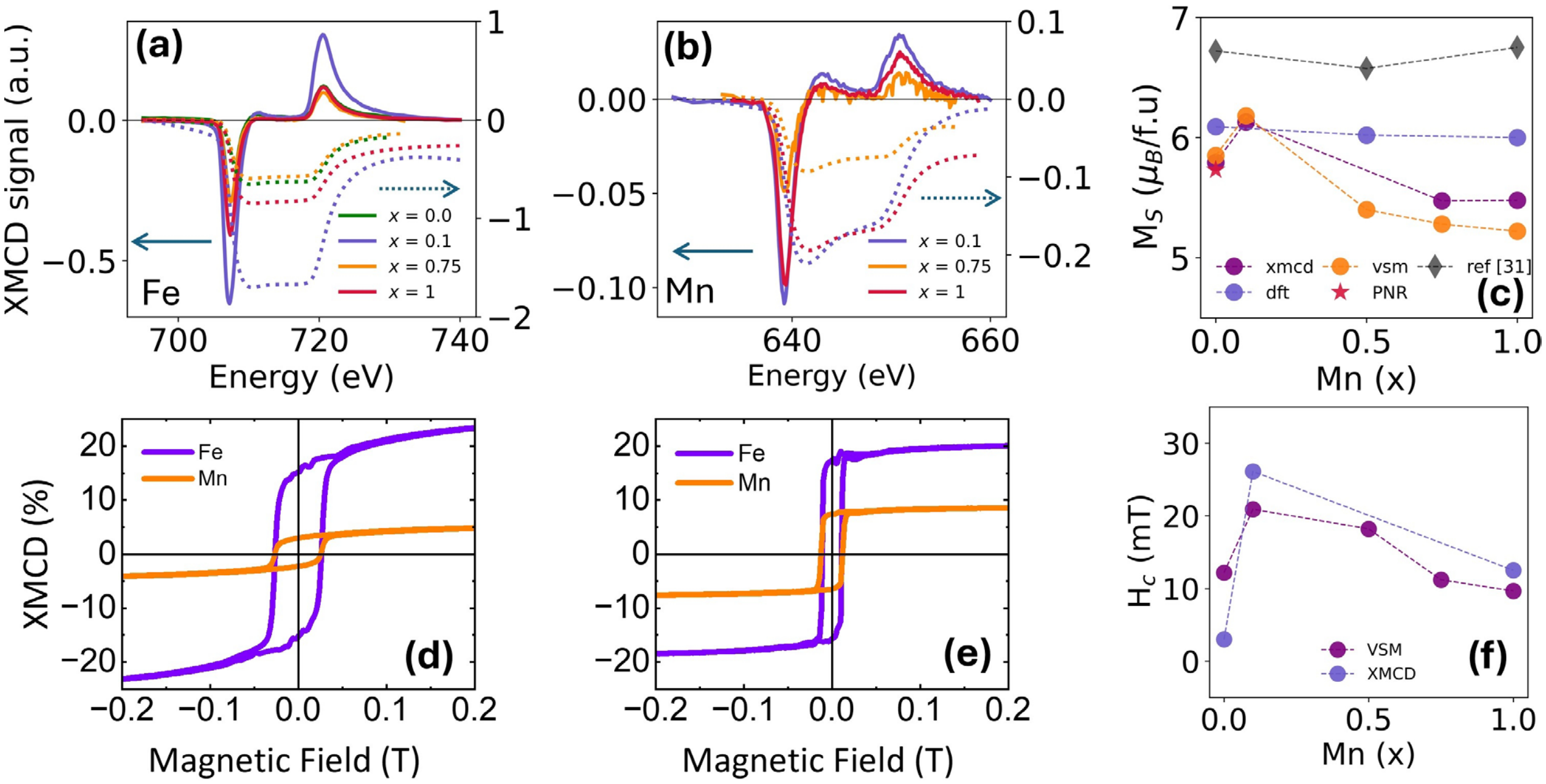

**FIG. 4.** XMCD difference signal of $Fe_{3-x}Mn_xSn_2$ thin films probed at the Fe (a) and Mn (b) $L_3$ and $L_2$ edges and their integral (dotted lines, right axis). (c) Total magnetic moment per formula unit of $Fe_{3-x}Mn_xSn_2$ obtained from several techniques. Field-dependent XMCD collected at fixed energy at the $L_3$ edges of Fe (707.5 eV) and Mn (639.3 eV) for (d) $Fe_{2.9}Mn_{0.1}Sn_2$ and (e) $Fe_2MnSn_2$. (f) Coercive field at differing Mn content obtained from magnetometry and XMCD data.

Beyond determining the moments in saturation, XMCD was also employed to investigate elementally resolved magnetization switching by measuring dichroic intensity at a fixed energy while sweeping the field through a hysteresis loop. Figure 4(d,e) shows XMCD hysteresis loops collected for $x = 0.1$ and 1 samples from -0.2 T to 0.2 T at the $L_3$ edge of Fe and Mn where the XMCD signal is maximum. The element specific $H_C$ of both Fe and Mn are equal (~27 mT in $Fe_{2.9}Mn_{0.1}Sn_2$ and ~12 mT in $Fe_2Mn_1Sn_2$), confirming that Fe and Mn moments are ferromagnetically aligned in the $Fe_{3-x}Mn_xSn_2$ lattice. Our measured values are consistent with other reports from $Fe_3Sn_2$ films, where $H_C$ ~ 10 mT was obtained from 80 nm thick films on Pt/Ru buffer layers.[37] A comparison between the coercive fields obtained from magnetometry and XMCD are shown in Fig. 4(f). Good agreement is obtained between the $H_C$ values measured with XMCD and lab-based magnetometry. These results suggest that Mn alloying does not significantly alter the magnetic anisotropy in thin films of $Fe_{3-x}Mn_xSn_2$ in which the easy axis lies within the *ab* plane as opposed to along the *c*-axis as in bulk; this change in easy axis is presumably brought on by the shape anisotropy of the thin film geometry.

## IV. DISCUSSION AND CONCLUSIONS

We have investigated the MBE deposition of Mn- and Co-alloyed $Fe_3Sn_2$ films on Co buffer layers. Mn-alloyed $Fe_{3-x}Mn_xSn_2$ ($0.1 \leq x \leq 1$) films were epitaxially synthesized and exhibit the same crystal structure as the kagome $Fe_3Sn_2$ phase. In contrast, we find that the Co-alloyed films phase-separate into distinct lateral regions consisting of Co-Fe and kagome (Co,Fe)Sn alloys. The latter has a stacking sequence similar to that of FeSn and CoSn with a single kagome layer separated by stanene. The observed phase separation has important implications for tuning the properties of $Fe_3Sn_2$ through electron doping, as it implies that Co substitution is not a viable strategy. Given the instability of $(Fe,Co)_3Sn_2$, alternative routes for electron doping will need to be established, such as charge transfer at $Fe_3Sn_2$-based interfaces, interfacial stabilization of $(Fe,Co)_3Sn_2$ as ultrathin layers in superlattices, or alloying $Fe_3Sn_2$ with Ni.

The single phase $Fe_{3-x}Mn_xSn_2$ films ($0 \leq x \leq 1$) exhibit a Fermi level shift with respect to the flat bands upon hole-doping, while retaining room temperature ferromagnetism. HAXPES measurements reveal an upward movement of the valence bands relative to the Fermi level and the measured spectra are well reproduced by DFT calculated VB spectra that incorporate photoemission cross sections. Our DFT calculations indicate that the Mn incorporation not only moves the Fermi level but also results in a redistribution of orbital character within the FBs. Through a combination of PNR and magnetometry, we confirmed a room temperature saturation magnetization of ~1.92 $\mu_B$/Fe in $Fe_3Sn_2$ films. The net magnetization decreases slightly with increased Mn concentration. The element contributions to this trend in net magnetization were determined through XMCD. A nonlinear evolution of coercive field obtained from XMCD and magnetometry data is observed as $x$ is increased from 0 to 1.0, which may hint at the role of disorder but detailed further studies are needed to better understand these trends. The broader conclusion of this work, that Mn enables the positions of $E_F$ and the FBs to be tuned without detriment to room temperature ferromagnetism, proves that Mn alloying is useful strategy to functional properties that arise from the flat bands such as anomalous Hall or Nernst effects.

## SUPPLEMENTARY MATERIAL

The supplementary materials provides reflection high energy electron diffraction patterns and in-plane x-ray diffraction data (Fig. S1), additional density functional theory calculations of electronic and magnetic properties (Figs. S2, S3), additional information on computed valence band spectra (Fig. S4) and sum rules analysis of x-ray magnetic circular dichroism data (Figs. S4-S8).

## ACKNOWLEDGEMENTS

Research primarily supported by the U.S. Department of Energy (DOE), Office of Science, Basic Energy Sciences (BES), under Award # DE-SC0024204 (neutron scattering, materials synthesis, x-ray scattering, x-ray spectroscopy analysis, magnetometry, density functional theory). A.X.G., S.S., and U.M.J. (photoelectron spectroscopy and analysis) acknowledge support from the U.S. Department of Energy, Office of Science, Office of Basic Energy Sciences, Materials Sciences and Engineering Division under Award No. DE-SC0024132. Work at Brown University was supported by the National Science Foundation under Grant No. 2429695. This work used resources of the Singh Center for Nanotechnology, which is supported by the NSF National Nanotechnology Coordinated Infrastructure Program under grant NNCI-2025608 and through the use of facilities supported by the University of Pennsylvania Materials Research Science and Engineering Center (MRSEC) DMR-2309043. A portion of this research used resources at the Spallation Neutron Source, a DOE Office of Science User Facility operated by the Oak Ridge National Laboratory; the beam time was allocated to MAGREF on proposal number IPTS- 32923. We would like to thank Hussein Hijazi at Rutgers University for performing timely RBS experiments. We also acknowledge use of the material characterization core facility at Drexel University.

## AUTHOR DECLARATIONS

### Conflict of Interest

The authors have no conflicts to disclose.

## DATA AVAILABILITY

The data that support this work are available from the corresponding author on responsible request.